# Bayesian Forecasting of Stock Returns on the JSE using Simultaneous Graphical Dynamic Linear Models

Nelson Kyakutwika[1*] and Bruce Bartlett[1,2]

[1]*Mathematics Division, Stellenbosch University, South Africa;* [2]*National Institute for Theoretical and Computational Sciences, South Africa*

Dedicated to the memory of Ronnie Becker

**ABSTRACT**

Cross-series dependencies are crucial in obtaining accurate forecasts when forecasting a multivariate time series. Simultaneous Graphical Dynamic Linear Models (SGDLMs) are Bayesian models that elegantly capture cross-series dependencies. This study forecasts returns of a 40-dimensional time series of stock data from the Johannesburg Stock Exchange (JSE) using SGDLMs. The SGDLM approach involves constructing a customised dynamic linear model (DLM) for each univariate time series. At each time point, the DLMs are recoupled using importance sampling and decoupled using mean-field variational Bayes. Our results suggest that SGDLMs forecast stock data on the JSE accurately and respond to market gyrations effectively.

## 1  Introduction

In the stock market, the price of any stock partly depends on the changes in the prices of the other stocks. Gruber and West (2016) introduced *Simultaneous Graphical Dynamic Linear Models (SGDLMs)* to address the need for capturing dependencies among time series while maintaining the flexibility of customising models at the level of individual time series. In their study, they used SGDLMs to forecast returns of a multivariate time series consisting of 400 stocks of the S&P 500 index, and their SGDLM outperformed the standard Wishart dynamic linear model. SGDLMs have been applied to portfolio management by Gruber and West (2017) and forecasting macroeconomic series by Xie (2021). Other studies on SGDLMs include Griveau-Billion and Calderhead (2019) and West (2020). Many multivariate models encounter difficulties when applied to high-dimensional time series (Gruber & West, 2016). Models becoming over-parameterised and the inability to scale up computationally, with increasingly high dimensions, are some of the difficulties (Gruber & West, 2016). SGDLMs are more parsimonious compared to traditional Bayesian multivariate models, e.g., the Wishart dynamic linear model, so they do not suffer heavily from these difficulties. However, not many studies have been conducted on these recently introduced models. Therefore, more research is needed, especially in elucidating further the structure of SGDLMs and the SGDLM algorithm. Capturing cross-series dependencies in high-dimensional time series usually presents a challenge of computing the many calculations involved; this is the basis upon which Griveau-Billion and Calderhead (2019), Gruber and West (2016, 2017), West (2020), and Xie (2021) used graphics processing units (GPUs) for their computations. Using the unconventional GPU-accelerated parallelisation greatly speeds up the computations in SGDLMs. Unlike the previous authors who worked with hundreds of time series (e.g., Gruber and West (2016, 2017) worked with 400 stocks and Griveau-Billion and Calderhead (2019) worked with 376 stocks), using the common central processing unit (CPU)-based computers should suffice when one has a smaller number of time series, for example, the 40 stocks of the current study. There is therefore a need to investigate the feasibility of applying the SGDLM in conjunction with CPU-based computers in situations when the dimension of the multivariate time series is low.

---

[*]Corresponding author. Email: nelsonkyakutwika@aims.ac.za

In the current study, we present the algorithm of the first version of SGDLMs in detail to make it easier for interested researchers to follow up. We apply a SGDLM to forecast daily log-returns of a multivariate time series consisting of 40 stocks from the South African Johannesburg Stock Exchange, which is a different market to the NYSE and NASDAQ where the SGDLM has exclusively been applied so far. We make a comparison between the forecasts of the dynamic linear model (DLM) and those of the SGDLM for a particular stock and investigate the effect of the number of simultaneous parents on forecast accuracy. We implement the SGDLM algorithm in Python and run the code on a local machine with CPU hardware.

This paper is structured as follows: Section 2 gives an overview of state space models and stochastic observational variance DLMs. Section 3 describes the structure of the SGDLM. Section 4 presents the SGDLM algorithm. Section 5 explains our SGDLM implementation, and results of the analyses are given in Section 6. Finally, Section 7 concludes the study. Supplementary material appears in the appendices.

## 2 State space models and stochastic variance DLMs

In this section we briefly review the notion of a Bayesian state space model for a time series. We also review a commonly used example of such a model (the stochastic variance dynamic linear model). The SGDLM is an example of this framework.

### 2.1 Bayesian state space models

Suppose we are trying to model a daily time series $y_t$ ($t = 0,1,2,...$). A *Bayesian state space model* (Petris et al., 2009, Sections 2.3 and 2.4) models $y_t$ as an explicit function of an evolving *state variable* $\Theta_t$:

$$\text{Observation equation:} \quad y_t = f(\Theta_t) + \eta_t \tag{1}$$

$$\text{Evolution equation:} \quad \Theta_{t+1} = g(\Theta_t) + \omega_t \tag{2}$$

Here $\eta_t$ and $\omega_t$ are stochastic noise terms, and $f$ and $g$ are explicitly known functions. Both the time series $y_t$ and the state variable $\Theta_t$ could be scalar or vector quantities, but we will write them in plain text for simplicity. (In our example later of the SGDLM algorithm, $y_t$ will be the vector of daily returns $y_{it}$ for all the stocks, while $\Theta_t$ will consist of regression coefficients $\phi_{it}$, coupling coefficients $\gamma_{ij,t}$ and precisions $\lambda_{it}$).

To understand how this works, imagine ourselves on the morning of day 0. The day has just begun, so data point $y_0^*$ is not yet known, as this is measured in the afternoon (say). But, we are given an initial probability distribution (the *prior*) $p(\Theta_0)$ for the state $\Theta_0$, allowing us to infer a conditional probability distribution $p(y_0|\Theta_0)$ using the observation equation (Equation (1)). To forecast the actual probability that $y_0$ will occur, unconditional on the state vector $\Theta_0$ (which, after all, consists of some internal 'hidden variables' that we are using in our model but which the public, who we are reporting to, does not care about), we must average over all $\Theta_0$ using the prior distribution for $\Theta_0$:

$$p(y_0) = \int p(y_0|\Theta_0)p(\Theta_0)d\Theta_0 \tag{3}$$

Now, time passes on day 0. We have our lunch, and then in the afternoon, we obtain our first actual data point $y_0^*$. We use this to *update our prior belief* about what the probability distribution for the state vector $\Theta_0$ was, given the data point $y_0^*$. This *posterior* distribution for $\Theta_0$ is written as $p(\Theta_0|y_0^*)$, and we compute it using *Bayes' formula*:

$$p(\Theta_0|y_0^*) = \frac{p(y_0^*|\Theta_0)p(\Theta_0)}{p(y_0^*)} \tag{4}$$

It is now time to *evolve* our state variable $\Theta_0$ to the following day, using the evolution equation (2). To do this, we first translate this evolution equation into a formula $p(\Theta_{t+1}|\Theta_t)$ for the probability that $\Theta_{t+1}$ will occur, given $\Theta_t$. Using this, we can evolve our *posterior probability distribution* $p(\Theta|y_0^*)$ for the state variable $\Theta_0$ at the close of day 0 into our *prior probability distribution* $p(\Theta_1|y_0^*)$ for $\Theta_1$ on the morning of day 1:

$$p(\Theta_1|y_0^*) = \int p(\Theta_1|\Theta_0)p(\Theta_0|y_0^*)\, d\Theta_0 \tag{5}$$

So now, on the morning of day 1, we are in a similar position to the one we were in on the morning of day 0. We have a prior belief about the state $\Theta_1$, from which we can forecast what the data point $y_1$ will be at the end of the day. And so the cycle continues.

We will follow Gruber and West (2016) and use the shorthand notation $\mathcal{D}_t$ for the information of all data points up to and including time $t$:

$$\mathcal{D}_t := \{y_0^*, y_1^*, \ldots, y_t^*\}$$

Note that $\mathcal{D}_{-1} = \emptyset$, i.e., no information. In this notation, we can summarize the Bayesian state space model process in the following diagram:

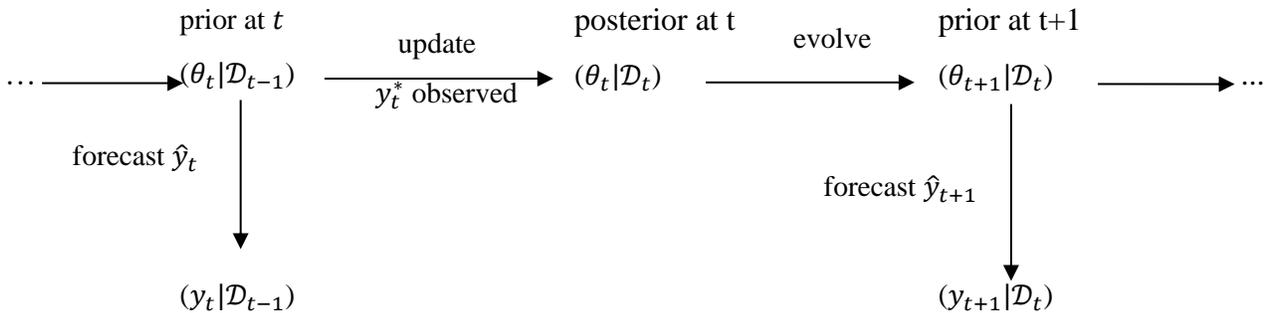

Given the initial prior distribution $p(\theta_1) = p(\theta_1|\mathcal{D}_0)$ for the state, our task is to compute explicit formulas for the posterior probability distribution $p(\theta_t|\mathcal{D}_t)$ and its evolution to the prior probability distribution $p(\theta_{t+1}|\mathcal{D}_t)$ for all $t \geq 0$. In practice, we would like to set up the model in such a way that these distributions have *conjugate forms* for all $t$ (the probability distribution remains fixed, only its parameters change). We will see an example of this in the next subsection.

## 2.2 The stochastic observational variance DLM with discounting

The *stochastic observational variance DLM* with discounting (Prado & West, 2010, Section 4.3.7) is an illustrative example of a state space model. It is also an important component in the SGDLM algorithm. For a comprehensive review, see Kyakutwika (2022).

In this model, we start by fixing an *observational variance discount factor* $\beta \in (0,1]$ and an *evolution variance discount factor* $\delta \in (0,1]$. The model is defined as follows.

Observation equation: $\quad y_t = \mathbf{F}_t^T \boldsymbol{\theta}_t + v_t \qquad v_t \sim N[0, \lambda_t^{-1}]$ (6)

Evolution equation: $\quad \boldsymbol{\theta}_t = \boldsymbol{\theta}_{t-1} + \boldsymbol{\omega}_t \qquad \boldsymbol{\omega}_t \sim N[\mathbf{0}, \mathbf{W}_t]$ (7)

Precision equation: $\quad \lambda_t = \dfrac{\eta_t}{\beta} \lambda_{t-1} \qquad \eta_t \sim \text{Be}\left[\dfrac{\beta n_{t-1}}{2}, \dfrac{(1-\beta) n_{t-1}}{2}\right]$ (8)

Initial prior at time 0: $\quad (\boldsymbol{\theta}_0, \lambda_0) \sim NG[\mathbf{a}_0, \mathbf{R}_0, r_0, c_0]$ (9)

Here,

- $y_t$ is a scalar and $\boldsymbol{\theta}_t$ is an $n$-dimensional column vector,
- the $n$-dimensional column vector $\mathbf{F}_t$ and the $n \times n$ covariance matrix $\mathbf{W}_t$ are explicitly known at each time $t$,
- $\lambda_t$ is the *observational precision* at time $t$ (i.e., the reciprocal of the variance of $y_t | \boldsymbol{\theta}_t$) and is itself a stochastic quantity,
- $\eta_t \in (0,1]$ is a *random shock* independent of $\lambda_{t-1}$, and governed by the beta distribution (the quantity $n_t$ will be known at time $t$, see below), and
- the vector $\mathbf{a}_0$, the matrix $\mathbf{R}_0$, and the scalars $r_0$ and $c_0$ are given.

By combining the state vector $\boldsymbol{\theta}_t$ and the observational precision $\lambda_t$ into a grand state vector $\boldsymbol{\Theta}_t$, we can view this model as an example of a state space model as in Section 2.1. The model (for instance the specification of the covariance matrices $\mathbf{W}_t$) is designed in such a way that the explicit solution for the prior and posterior distributions remains of the same form (i.e., conjugate) for all $t$:

$$\begin{aligned} \text{prior at time } t: \quad & (\theta_t, \lambda_t | \mathcal{D}_{t-1}) \sim NG(\mathbf{a}_t, \mathbf{R}_t, r_t, c_t) \\ \text{posterior at time } t: \quad & (\theta_t, \lambda_t | \mathcal{D}_t) \sim NG(\mathbf{m}_t, \mathbf{C}_t, n_t, s_t) \end{aligned}$$

Therefore, the complete solution of the model may be specified by giving explicit recursive formulas (the *Kalman filter*) for the above parameters, which are as follows. (See Kyakutwika (2022) and references therein for a derivation.)

*2.2.1 Updating equations*

Given the data point $y_t^*$, we update the prior at time $t$ to the posterior at time $t$ as follows. We firstly compute:

| | |
|---|---|
| 1-step ahead forecast error: | $e_t = y_t^* - \mathbf{F}_t^T \mathbf{a}_t$ |
| 1-step ahead forecast variance factor: | $q_t = c_t + \mathbf{F}_t^T \mathbf{R}_t \mathbf{F}_t$ |
| Adaptive coefficient vector: | $\mathbf{A}_t = \dfrac{1}{q_t} \mathbf{R}_t \mathbf{F}_t$ |
| Volatility update factor: | $z_t = \left(r_t + \dfrac{e_t^2}{q_t}\right) \dfrac{1}{r_t + 1}$ |

Then we set:

| | |
|---|---|
| Posterior mean vector: | $\mathbf{m}_t = \mathbf{a}_t + e_t \mathbf{A}_t$ |
| Posterior covariance matrix factor: | $\mathbf{C}_t = z_t (\mathbf{R}_t - q_t \mathbf{A}_t \mathbf{A}_t^T)$ |
| Posterior degrees of freedom: | $n_t = r_t + 1$ |
| Posterior observational variance estimate: | $s_t = z_t c_t$ |

*2.2.2 Evolution equations*

We evolve the posterior at time $t$ to the prior at time $t+1$ as follows (note the use of the discount factors):

$$\mathbf{a}_{t+1} = \mathbf{m}_t, \quad \mathbf{R}_{t+1} = \frac{1}{\delta} \mathbf{C}_t, \quad r_{t+1} = \beta n_t, \quad c_{t+1} = s_t \tag{10}$$

*2.2.3 Evolution equations with block discounting*

For later use we record a version of the evolution equations, called *block discounting* (Prado and West 2010, sec. 4.3.7), applicable when the state vector $\boldsymbol{\theta}_t$ naturally partitions into two vectors,

$$\boldsymbol{\theta}_t = \begin{bmatrix} \boldsymbol{\theta}_t^1 \\ \boldsymbol{\theta}_t^2 \end{bmatrix}.$$

We aim to have separate evolution variance discount factors, $\delta_1$ and $\delta_2$, for each part of the state vectors. Given that the matrix $\mathbf{R}_0$ in the initial prior is typically block-diagonal, we strive to preserve this block-diagonal structure for $\mathbf{R}_t$ at all times $t$. Consequently, we disregard the non-block-diagonal components of $\mathbf{C}_t$ in the updated evolution equation for $\mathbf{R}_{t+1}$. Thus, with block-discounting, the evolution equations mirror Equation (10), but with a modified formula for $\mathbf{R}_{t+1}$:

$$\mathbf{a}_{t+1} = \mathbf{m}_t, \quad \mathbf{R}_{t+1} = \begin{pmatrix} \frac{1}{\delta_1} \mathbf{C}_{t[1,1]} & \cdots \\ \cdots & \frac{1}{\delta_2} \mathbf{C}_t[2:,2:] \end{pmatrix}, \quad r_{t+1} = \beta n_t, \quad c_{t+1} = s_t \quad (11)$$

## 3  The SGDLM for stock prices

Suppose that there are $m$ stocks on the stock exchange. The SGDLM postulates that the closing return $y_{it}$ of stock $i$ on day $t$ is linearly related with the closing returns

$$\{y_{jt},\ j \in \mathrm{sp}(i)\}$$

of a set of *other stocks* $\mathrm{sp}(i) \subset \{1,2,\ldots,m\} \setminus \{i\}$ on the same day $t$, the *simultaneous parents* of stock $i$. (In this paper, we follow Gruber and West 2016 and stipulate that all stocks have the same number $k$ of simultaneous parents, i.e., $|\mathrm{sp}(i)| = k$ for all $i$). Specifically, the model postulates that

$$y_{it} = \phi_{it} + \sum_{j \in \mathrm{sp}(i)} \gamma_{ij,t}\, y_{j,t} + \nu_{i,t}, \quad \nu_{it} \sim \mathrm{N}[0, \lambda_{it}^{-1}] \quad (12)$$

for some *regression coefficients* $\phi_{i,t}$ and *coupling coefficients* $\gamma_{ij,t}$.

To describe the model further, it helps to understand it from two complementary viewpoints.

### 3.1  The SGDLM as a collection of coupled DLMs

If we write

$$\mathbf{F}_{it} = \begin{bmatrix} 1 \\ \gamma_{ij_1,t} \\ \vdots \\ \gamma_{ij_k,t} \end{bmatrix} \quad \text{and} \quad \boldsymbol{\theta}_{it} = \begin{bmatrix} \phi_{it} \\ y_{j_1 t} \\ \vdots \\ y_{j_k t} \end{bmatrix},$$

then we can embed Equation (12) into a complete specification of the SGDLM as a collection of $m$ coupled stochastic observational variance DLMs with discount factors, as in Equations (6) to (9):

$$y_{it} = F_{it}^T \theta_{it} + v_{it}, \qquad v_{it} \sim N[0, \lambda_{it}^{-1}] \qquad (13)$$

$$\theta_{it} = \theta_{i,t-1} + \omega_{it}, \qquad \omega_{it} \sim N[0, W_{it}] \qquad (14)$$

$$\lambda_{it} = \frac{\eta_{it}}{\beta} \lambda_{i,t-1}, \qquad \eta_t \sim Be\left[\frac{\beta n_{i,t-1}}{2}, \frac{(1-\beta) n_{i,t-1}}{2}\right] \qquad (15)$$

$$(\theta_{i0}, \lambda_{i0}) \sim NG[a_{i0}, R_{i0}, r_{i0}, c_{i0}] \qquad (16)$$

To understand precisely what this collection of 'coupled DLMs' means, we need to view these equations from a more global point of view.

### 3.2   The SGDLM as a nonlinear state space model

The coupling between time series in Equation (13) means that a SGDLM is in fact a nonlinear state space model. To see this, collect all the regression coefficients $\phi_{it}$ into an $m$-dimensional vector $\boldsymbol{\phi}_t$, and all the coupling coefficients $\gamma_{ij,t}$ into the matrix

$$\Gamma_t = \begin{pmatrix} 0 & \gamma_{12,t} & \gamma_{13,t} & \cdots & \gamma_{1m,t} \\ \gamma_{21,t} & 0 & \gamma_{23,t} & \cdots & \gamma_{2m,t} \\ \gamma_{31,t} & \gamma_{3,2,t} & 0 & \cdots & \gamma_{3m,t} \\ \vdots & \vdots & \vdots & \ddots & \vdots \\ \gamma_{m1,t} & \gamma_{m2,t} & \gamma_{m3,t} & \cdots & 0 \end{pmatrix}.$$

Note that $\Gamma_t$ is a sparse matrix (most of the entries are zero). Using this matrix notation, we can rewrite Equation (13) in column-vector form,

$$\mathbf{y}_t = \boldsymbol{\phi}_t + \Gamma_t \mathbf{y}_t + \mathbf{v}_t,$$

where $\mathbf{v}_t = (v_{1t}, \dots, v_{mt})^T$. Note that $\mathbf{y}_t$ appears on both sides of this equation. But we can solve for $\mathbf{y}_t$ explicitly as

$$\mathbf{y}_t = A_t(\boldsymbol{\phi}_t + \mathbf{v}_t)$$

where $A_t = (I - \Gamma_t)^{-1}$. Therefore, the probability distribution for $\mathbf{y}_t$ is

$$(\mathbf{y}_t | \boldsymbol{\theta}_t, \boldsymbol{\lambda}_t) \sim N[A_t \boldsymbol{\phi}_t, A_t \Lambda_t^{-1} A_t^T], \qquad (17)$$

where

$$\boldsymbol{\lambda}_t = (\lambda_{1t}, \lambda_{2t}, \dots, \lambda_{mt})^T$$

is the vector of observational precisions, $\Lambda_t = \text{diag}(\lambda_{1t}, \ldots, \lambda_{mt})$ is the corresponding diagonal matrix, and $\boldsymbol{\theta}_t$ is shorthand for the collection of all 'individual state vectors' $\boldsymbol{\theta}_{it}$.

In fact, we can think of all the individual state vectors $\boldsymbol{\theta}_{it}$ and observational precisions $\lambda_{it}$ as together forming a single 'grand state vector' $\boldsymbol{\Theta}_t$ as we did in the stochastic variance DLM from Section 2.2. From this perspective, we view Equation (17) as being the observation equation expressing the SGDLM as a nonlinear state-space model for the stock market, where the returns vector $\mathbf{y}_t$ is computed nonlinearly from a state vector $\boldsymbol{\Theta}_t$. The state of the market at time $t$ is encoded by three sets of parameters: the $m$ regression coefficients $\phi_{i,t}$ (the 'internal state' of stock $i$ at time t), the $mk$ coupling coefficients $\gamma_{ij,t}$ (the strength with which stock $i$ is coupled to stock $j$ at time $t$), and the $m$ observational precisions $\lambda_{it}$ (the precision at which the returns vector can be computed from the state vector). The nonlinearity arises because the inverse in $(\mathbf{I} - \boldsymbol{\Gamma}_t)^{-1}$ gives a degree $m$ polynomial dependence of $\mathbf{y}_t$ on the coupling coefficients $\gamma_{ij,t}$.

This nonlinear dependence of the returns vector $\mathbf{y}_t$ on the state variables makes it impossible to analytically evaluate the integral in Equation (5) to give a formula for evolving the state vector to the next day. Therefore, we will use a mean-field Monte Carlo approach (see Section 4), where each stock price $y_{it}$ is thought of as a univariate variable being acted on by the `background field' of the other stock prices, but not directly affecting them (much as to a first approximation, the dynamics of the moon is treated as moving in the `fixed' background field of the earth and the other planets).

### 3.3 *Factorization properties of the model*

Despite the complexity in evolving the state variables, updating them upon the arrival of a data point of a returns vector $\mathbf{y}_t^*$ can be done exactly due to the model's factorization properties. The full multivariate normal probability distribution (Equation (17)) for $(\mathbf{y}_t | \boldsymbol{\theta}_t, \boldsymbol{\lambda}_t)$ factorizes as a multivariate determinant multiplied by a product of univariate DLM distributions (see Kyakutwika (2022), Appendix C for a proof):

$$p(\mathbf{y}_t | \boldsymbol{\theta}_t, \boldsymbol{\lambda}_t) = |\mathbf{I} - \boldsymbol{\Gamma}_t| \prod_{i=1}^{m} p(y_i | \theta_{it}, \lambda_{it}) \qquad (18)$$

Here, $p(y_{it}|\boldsymbol{\theta}_{it},\lambda_{it})$ is really shorthand for $p(y_{it}|\boldsymbol{\phi}_{it},\{\gamma_{ij,t},y_{jt}\}_{j\in\text{sp}(i)})$, that is, it is the probability for obtaining $y_{it}$ given knowledge of the *other* $y_{jt}$ values, as in Equation (12).

This implies, by Bayes' rule, a corresponding factorization formula for how to update our knowledge of the state $(\boldsymbol{\theta}_t, \boldsymbol{\lambda}_t)$ when the data point of a returns vector $\mathbf{y}_t^*$ arrives:

$$p(\boldsymbol{\theta}_t, \boldsymbol{\lambda}_t | \mathbf{y}_t^*) \propto p(\mathbf{y}^* | \boldsymbol{\theta}_t, \boldsymbol{\lambda}_t) p(\boldsymbol{\theta}_t, \boldsymbol{\lambda}_t)$$

$$= |\mathbf{I} - \boldsymbol{\Gamma}_t| \prod_{i=1}^{m} p(y_i^* | \theta_{it}, \lambda_{it}) p(\theta_{it}, \lambda_{it})$$

$$\therefore p(\boldsymbol{\theta}_t, \boldsymbol{\lambda}_t | \mathbf{y}_t^*) \propto |\mathbf{I} - \boldsymbol{\Gamma}_t| \prod_{i=1}^{m} p(\theta_{it}, \lambda_{it} | y_{it}^*)$$

$$\therefore p(\boldsymbol{\theta}_t, \boldsymbol{\lambda}_t | \mathbf{y}_t^*) \propto |\mathbf{I} - \boldsymbol{\Gamma}_t| \prod_{i=1}^{m} p(\theta_{it}, \lambda_{it} | y_{it}^*) \quad (19)$$

Here, the first line is Bayes' rule, the second line is the factorization (Equation (18)) together with the assumption that $(\boldsymbol{\theta}_t, \boldsymbol{\lambda}_t)$ is a product distribution *before* the data point $\mathbf{y}_t^*$ arrives, and the third line is Bayes' rule in reverse. Equation (19) is useful as it tells us that the exact Bayesian updated multivariate probability in the model is a product of a multivariate determinant with the product of the updated univariate DLM probabilities (which we have concrete formulas for, via the Kalman filter in Section 2.2.1).

## 4 The SGDLM algorithm

Having explained the SGDLM, we can now clearly summarise the algorithm for its numerical implementation. See also Griveau-Billion and Calderhead (2019), Gruber and West (2016), and Gruber and West (2017).

### 0. Fix initial parameters

First fix the number of simultaneous parents, $k$. Then fix an observational variance discount factors $\beta \in (0,1]$, and two evolution variance discount factors $\delta_\phi, \delta_\gamma \in (0,1]$, catering for the regression coefficient $\phi_i$ and coupling coefficients $\gamma_{ij}$ for all individual state vectors $\boldsymbol{\theta}_{it}$ respectively. Finally, the initial parameters $(\mathbf{a}_{i0}, \mathbf{R}_{i0}, r_{i0}, c_{i0}), i = 1{:}m$ must be chosen. Each $\mathbf{a}_{i0}$ is a $(k+1)$-dimensional vector and each $\mathbf{R}_{i0}$ is a $(k+1)$-dimensional square matrix.

### 1. Initial priors at time $t$

At time $t$, we have decoupled priors for the state variables:

$$p(\boldsymbol{\theta}_t, \boldsymbol{\lambda}_t | \mathcal{D}_{t-1}) = \prod_i NG\left[\boldsymbol{a}_{it}, \boldsymbol{R}_{it}, r_{it}, c_{it}\right]$$

## 2. Make predictions at time $t$

To make predictions for $\mathbf{y}_t$, draw $K$ samples $\{(\boldsymbol{\theta}_t^r, \boldsymbol{\lambda}_t^r), r = 1:K\}$ from the decoupled distribution above (this step can be done efficiently using parallel computations). Then use Equation (17) to obtain a simulation sample of $K$ values of $\mathbf{y}_t$.

## 3. Update prior to posterior

After the data point $\mathbf{y}_t^*$ of returns arrives, the exact updated posterior is computed using Equation (19),

$$p_{\text{exact}}(\boldsymbol{\theta}_t, \boldsymbol{\lambda}_t | \mathcal{D}_t) \propto |\mathbf{I} - \boldsymbol{\Gamma}_t| \prod_i NG\left[\tilde{\mathbf{m}}_{it}, \tilde{\mathbf{C}}_{it}, \tilde{r}_{it}, \tilde{c}_{it}\right]$$

where the parameters of the normal-gamma factors have been updated using the standard Kalman filter updating equations (Section 2.2.1) for a stochastic observational variance DLM.

## 4. Approximate the posterior as a product

To prepare to evolve the posterior forward, we first need to approximate it as a product. This is done in two steps.

### a. Approximating the exact posterior as a discrete distribution (recouple)

We firstly approximate the continuous posterior $p_{\text{exact}}$ by a discrete Monte-Carlo posterior

$$p_{\text{MC}} = \sum_{n=1}^{N} w_t^n \left(\boldsymbol{\theta}_t^n, \boldsymbol{\lambda}_t^n\right)$$

by drawing an *importance sample* $\{(\boldsymbol{\theta}_t^n, \boldsymbol{\lambda}_t^n), n = 1:N\}$ efficiently in parallel from the naïve factorized posterior (which leaves out the determinant):

$$p_{\text{naïve}}(\boldsymbol{\theta}_t, \boldsymbol{\lambda}_t | \mathcal{D}_t) = \prod_i NG\left[\tilde{\mathbf{m}}_{it}, \tilde{\mathbf{C}}_{it}, \tilde{r}_{it}, \tilde{c}_{it}\right]$$

We calculate the weight $w_t^n$ of each sample $(\boldsymbol{\theta}_t^n, \boldsymbol{\lambda}_t^n)$ from the determinant $|\mathbf{I} - \boldsymbol{\Gamma}_t|$, and then normalize the weights so that they add to 1.

## b. Approximating the exact posterior as a product (decouple)

We then apply a mean-field approach and seek to find the best approximation $q$ to the exact posterior as a product:

$$p_{\text{exact}} \approx q = \prod_i \text{NG}\,[\mathbf{m}_{it}, \mathbf{C}_{it}, r_{it}, c_{it}] \qquad (20)$$

The parameters $\mathbf{m}_{it}, \mathbf{C}_{it}, r_{it}, c_{it}$ are found by a variational Bayes approach, which minimises the Kullback-Leibler (KL) divergence between the exact distribution (for which we use $p_{\text{MC}}$ as a computational substitute) and the candidate product distribution. For precise formulas, see Appendix C. We can approximate the KL divergence of $q$ from $p_{MC}$ using

$$KL(p_{\text{MC}}||q) \approx \sum_{i=1}^{N} w_{it}^n \ln(N w_{it}^n)$$

and use this to check the accuracy of our simulation.

## 5. Evolution to time $t+1$

We then evolve each factor in the product posterior $q$ from Equation (20) forward one time step, using the evolution equations (11) for a stochastic variance DLM with block-discounting. This gives the prior at time $t+1$.

## 5   Implementation of the SGDLM algorithm

### 5.1   Data set

Our data set is the *daily log-returns* of 40 JSE stocks that were selected from the Top 100 JSE index. The stocks, and their sector categorisations, are given in Appendix A. We downloaded the daily closing prices of all the stocks from Yahoo! Finance for the period 01/01/2014 to 30/06/2022 and calculated the log-returns. For the entire period, the total number of observations (the daily-log returns) is 2161 for each stock.

We implemented the SGLDM algorithm in Python. The data set was divided into the training set and the test set. The training set was further divided into two subsets, one for selecting simultaneous parents and the other for selecting discount factors. The data from 01/01/2014 to 31/12/2016 (782 observations) was used to select simultaneous parents; the data from 01/01/2017 to 31/12/2018 (506 observations) was used to select discount factors and obtain starting values for the test data analysis; and the data from 01/01/2019 to 30/06/2022 (873 observations) is the test set. Therefore, we divided the implementation into three phases: phase 1 (selection of simultaneous parents); phase 2 (selection of discount factors and initial priors for phase 3); and phase 3 (stock return forecasting).

## 5.2  Selection of simultaneous parents

Here, we ran the Kalman filter equations, stock by stock, for each of the 40 stocks. This phase entails implementing steps 0, 1, 3, and 5 of the SGDLM algorithm. Note that step 4 is not included in this phase, rather the analysis involves simply running the Kalman filter for each of the decoupled series. We adopted the initial priors of Gruber and West (2016) for this phase; these are $\mathbf{a}_{i0} = (0, \ldots, 0)^T$, $\mathbf{R}_{i0} = \text{diag}(0.0001, 0.01, \ldots, 0.01)$, $r_{i0} = 5$, and $c_{i0} = 0.001$, where $\mathbf{a}_{i0}$ is a $40 \times 1$ vector and $\mathbf{R}_{i0}$ is a $40 \times 40$ diagonal matrix whose first diagonal entry is $0.0001$ but the rest are $0.01$. All the stocks used the same initial prior. In this phase, every stock had all the remaining 39 stocks as simultaneous parents. We specified the evolution variance $\mathbf{W}_{it}$ using two discount factors, via standard block discounting. Therefore, we defined $\mathbf{W}_{it}$ as follows,

$$\mathbf{W}_{it} = \begin{pmatrix} \frac{1-\delta_\phi}{\delta_\phi} \tilde{\mathbf{C}}_{it}[1,1] & \mathbf{0} \\ \mathbf{0} & \frac{1-\delta_\gamma}{\delta_\gamma} \tilde{\mathbf{C}}_{it}[2:,2:] \end{pmatrix},$$

where $\tilde{\mathbf{C}}_{it}[1,1]$ is the first diagonal element of $\tilde{\mathbf{C}}_{it}$ and $\tilde{\mathbf{C}}_{it}[2:,2:]$ is the matrix that remains after deleting the first row and the first column of $\tilde{\mathbf{C}}_{it}$. The upper-left block $\tilde{\mathbf{C}}_{it}[1,1]$ is the local-level component whereas the lower-right block $\tilde{\mathbf{C}}_{it}[2:,2:]$ is the simultaneous parents component.

With the current dimension of 40 stocks, we found out that the SGDLM analysis is most accurate if every stock has just one simultaneous parent (see Section 6.5). On the last day of the period 01/01/2014 to 31/12/2016, for each stock $i$, we chose a stock's simultaneous parent from the other 39 stocks, depending on the absolute values of the posterior means of the vector $\boldsymbol{\gamma}_{it}$. As it can be seen from Equation (12), the numbers $\gamma_{ijt}$ are a measure of the effect of each of the other 39 stocks on stock $i$ (effect size). The simultaneous parent to stock $i$ is the stock that corresponds to the biggest effect size.

In Table 1, we give some selected stocks together with their simultaneous parents as generated by our analysis. We have underlined the simultaneous parent if it falls in the same sector with the stock it predicts. We notice that some of the simultaneous parents fall in the same sector with the stock being predicted – this causal relationship is expected. However, in some situations, the predictor and the stock being predicted fall in different sectors. This is still fine because dependencies in an economy can cut across sectors.

Table 1: Simultaneous parents for some selected stocks.

| **Stock** | **Simultaneous parent** |
|---|---|
| FirstRand Limited | <u>Standard Bank Group</u> |
| Standard Bank Group | <u>Nedbank Group Limited</u> |
| MTN Group Limited | ABSA Group Limited |
| British American Tobacco | Investec Limited |
| Compagnie Fin Richemont | Mr Price Group |
| Naspers | Aspen Pharmacare Holdings Limited |
| Truworths International Limited | <u>Mr Price Group</u> |
| Shoprite Holdings Limited | Nedbank Group Limited |
| Glencore plc | <u>Anglo American plc</u> |
| Anglo American plc | Clicks Group Limited |

### 5.3  *Selection of discount factors and obtaining initial priors for phase* 3

The discount factors to be selected are (i) $\beta$ (for learning the stochastic variance) and (ii) $\delta_\phi$ and $\delta_\gamma$ (for specifying the evolution variance). Phase 2 involves running all the steps of the SGDLM algorithm save for step 2. The discount factors are selected using the decoupled DLMs by maximising the log-likelihood function series by series, e.g., Prado and West, 2010, Section 4.3.6. The initial priors are like those used in phase 1, but since the analysis of the current phase uses only one simultaneous parent, $\mathbf{a}_{i0} = (0,0)^T$ and $\mathbf{R}_{i0} = \text{diag}(0.0001, 0.01)$. The evolution variance is now of the form

$$\mathbf{W}_{it} = \begin{pmatrix} \frac{1-\delta_\phi}{\delta_\phi} c_{1,1,i,t} & 0 \\ 0 & \frac{1-\delta_\gamma}{\delta_\gamma} c_{2,2,i,t} \end{pmatrix},$$

where the scalars $c_{1,1,i,t}$ and $c_{2,2,i,t}$ are the diagonal entries of the covariance matrix $\mathbf{C}_{it}$ of the exact posterior obtained in step 4 of the algorithm.

We explain how we determined $\delta_\gamma$. The other two discount factors were determined in a similar way. The predictive distribution of each of the decoupled time series is represented as $(y_{it}|\mathcal{D}_{i,t-1}) \sim T_{r_{it}}[f_{it}, q_{it}]$. The log-likelihood for stock $i$, for the period $t = 783$ to $t = 1288$, is then defined as

$$\log_e p(y_{i,783:1288}|\mathcal{D}_{i,782}, \delta_{\gamma i}) = \log_e \prod_{t=783}^{1288} p(y_{it}|\mathcal{D}_{i,t-1}, \delta_{\gamma i}) \tag{21}$$

We kept $\beta_i$ and $\delta_{\phi i}$ constant and varied $\delta_{\gamma i}$. (The values of $\beta_i$ and $\delta_{\phi i}$ were uniform across all stocks.) For different values of $\delta_{\gamma i}$, we obtained the log-likelihood using Equation (21) at the level of individual stocks. For the running example, after inspection, we observed that most of the values of $\delta_{\gamma i}$ were on the interval [0.859, 0.999]. So, we varied $\delta_{\gamma i}$ on this interval for each stock.

Table 2: Log-likelihood values at different values of $\delta_{\gamma i}$.

| $\delta_{\gamma i}$ | Log-likelihood | |
|---|---|---|
|  | Standard Bank | MTN Group |
| 0.859 | 1480 | 1280 |
| 0.894 | **1495** | 1285 |
| 0.929 | 1480 | 1288 |
| 0.964 | 1471 | **1292** |
| 0.999 | 1404 | 1287 |

In Table 2, we show the log-likelihood values that correspond to the different values of $\delta_{\gamma i}$ for two companies, Standard Bank and MTN Group. The optimal value of the discount factor is the one that corresponds to the maximum log-likelihood. Therefore, for Standard Bank, $\delta_\gamma = 0.894$ and for MTN Group, $\delta_\gamma = 0.964$. We obtained the value of $\delta_\gamma$ for the remaining stocks in a similar way and computed the average across all stocks. This average now serves as the discount factor for each stock. With the current example, this average is 0.953. Thus, $\delta_\gamma = 0.953$, which is taken uniform across all stocks. In a similar way, by keeping $\delta_\gamma$ and $\beta$ constant, we obtained $\delta_\phi$ as 0.993. And by keeping $\delta_\gamma$ and $\delta_\phi$ constant, we obtained $\beta$ as 0.922. Then, using these optimal values of the discount factors and the same initial priors, steps 0, 1, 3, 4, and 5 were re-run to obtain starting values for phase 3. The size of the importance sample ($N$) in this phase was kept at $N = 2,000$.

## 5.4 Stock return forecasting

In the test phase, we ran the all the six steps of the SGDLM algorithm for the last three and half years of our study period. This phase used the discount factors and initial priors obtained in phase 2. The analysis used $K = N = 2{,}000$.

## 6 Results from the test data analysis

### 6.1 Coverage of prediction intervals

In the context of time series forecasting, a *prediction interval*, aka *forecast interval*, is the interval which is constructed around the forecast, within which the observation is expected to lie with a specified probability (Hyndman and Athanasopoulos, 2018, Section 3.5). For example, the 95% prediction interval $[a, b]$ (constructed around the forecast $\hat{y}_{it}$) means that, according to the predicting model, there is a 95% probability that the observation $y_{it}$ will lie within the interval $[a, b]$. We calculated prediction intervals at the level of individual stocks using the large sample formula (e.g., Ramachandran and Tsokos, 2014, Section 14.5)

$$\hat{y}_{it} \pm z_{\alpha/2}\sqrt{\Sigma_{i,i,t}}\sqrt{1 + \frac{1}{K}},$$

where: $\hat{y}_{it}$ is the *forecast* that corresponds to the observation $y_{it}$; $z_{\alpha/2}$ is the *critical value* of the standard normal distribution and $1 - \alpha$ is the *degree of confidence*; $\Sigma_{i,i,t}$, the $i^{\text{th}}$ diagonal element of the covariance matrix $\mathbf{\Sigma}_t = \mathbf{A}_t \mathbf{\Lambda}_t^{-1} \mathbf{A}_t^T$, is the *variance* of $\hat{y}_{it}$; and $K$ is the forecasting *simulation sample size*.

For a perfect model, empirical coverage is equal to theoretical coverage. But because of the noise in the data and sometimes errors in the model, empirical coverage is not always equal to theoretical coverage. In practice, outputs of models portray under-coverage or over-coverage of intervals. The closer the output of the model to the theoretical coverage, the more accurate is the model. Over-coverage is preferred to under-coverage of the same magnitude because the former is coverage that is more than what is enough. We calculated the interval coverages at different levels of confidence for all the stocks throughout the entire test period. From $\mathbf{y}_t \sim N[\mathbf{A}_t \mathbf{\phi}_t, \mathbf{\Sigma}_t]$, each $y_{it} \sim N[\hat{y}_{it}, \Sigma_{i,i,t}]$. Using this result, we calculated the prediction intervals for all the stocks at the following levels of confidence: 99% $(z_{\alpha/2} = 2.58)$, 95% $(z_{\alpha/2} = 1.96)$, 90% $(z_{\alpha/2} = 1.64)$, 80% $(z_{\alpha/2} = 1.28)$, 50% $(z_{\alpha/2} = 0.67)$, 20% $(z_{\alpha/2} = 0.25)$, and 10% ($z_{\alpha/2} = 0.13$). In Table 3, we give the average interval coverages across all stocks and the interval coverages for eight of the forty stocks, for the entire test period. We also include the aggregate interval coverages of Gruber and West (2016) as benchmark values.

Table 3: Average interval coverage across all stocks/aggregate interval coverage and interval coverage for some selected stocks, for the entire test period.

| Prediction interval (%) | 99 | 95 | 90 | 80 | 50 | 20 | 10 |
|---|---|---|---|---|---|---|---|
| **Aggregate interval coverage** | | | | | | | |
| Coverage (%) | 98.4 | 95.7 | 92.6 | 86.0 | 60.4 | 26.5 | 14.4 |
| **Benchmark aggregate interval coverage** | | | | | | | |
| Coverage (%) | 98.4 | 95.6 | 92.4 | 85.5 | 59.7 | 27.2 | 14.4 |
| **Standard Bank Group** | | | | | | | |
| Coverage (%) | 98.9 | 95.5 | 92.3 | 85.9 | 60.5 | 27.1 | 14.2 |
| **FirstRand Limited** | | | | | | | |
| Coverage (%) | 99.2 | 96.3 | 93.4 | 86.3 | 59.8 | 25.0 | 14.0 |
| **Glencore plc** | | | | | | | |
| Coverage (%) | 99.2 | 96.0 | 92.1 | 84.4 | 58.2 | 22.6 | 11.5 |
| **Anglo American plc** | | | | | | | |
| Coverage (%) | 98.7 | 95.0 | 92.6 | 85.8 | 58.8 | 24.5 | 12.5 |
| **British American Tobacco** | | | | | | | |
| Coverage (%) | 98.5 | 95.0 | 92.1 | 84.3 | 59.6 | 27.6 | 14.1 |
| **MTN Group Limited** | | | | | | | |
| Coverage (%) | 97.7 | 96.0 | 94.3 | 86.8 | 63.0 | 28.3 | 16.0 |
| **Naspers** | | | | | | | |
| Coverage (%) | 97.9 | 95.0 | 90.8 | 84.8 | 61.1 | 26.6 | 14.5 |
| **Shoprite Holdings Limited** | | | | | | | |
| Coverage (%) | 97.7 | 95.5 | 93.1 | 87.5 | 62.0 | 27.8 | 15.9 |

According to Table 3, the aggregate interval coverages from 10% to 95% are bigger than the theoretical values. Nevertheless, these interval coverages are more precise compared to outputs of other multivariate models, e.g., see Gruber and West (2016). The 99% prediction interval is under-estimated in both the aggregate analysis and for most of the individual stocks, but empirical coverage remains close to the nominal one. We also observe that the interval coverages for each of the eight stocks are like those of the aggregate analysis. These realised SGDLM interval coverages are therefore literally tolerable. Finally, our aggregate interval coverage estimates compare nicely with those of the benchmark study (Gruber and West (2016)).

## 6.2 Comparison between the SGDLM and the DLM

In addition to predicting the daily log-returns using the SGDLM, we independently predicted the returns of each of the eight stocks in Table 3 using the stochastic volatility local-level DLM. In the DLM analysis, we partitioned the data in a way that is like that of the SGDLM analysis. We used the data from 01/01/2017 to 31/12/2018 (506 observations) to select the discount factors $\beta_i$ and $\delta_i$, and to get the initial values for the testing phase. The initial values of this training period were taken as $a = 0$, $R = 0.0001$, $c = 0.001$, and $r = 5$. The test data is from 01/01/2019 to 30/06/2022 (873 observations). We never used the data from 01/01/2014 to 31/12/2016 (782 observations) as this was purposely for selecting simultaneous parents in the SGDLM case.

If, for example, we are interested in forecasting the price of Standard Bank on a daily basis, we can use either the SGDLM where Standard Bank will be modelled together with other stocks or the DLM that will focus on Standard Bank alone. So, in the SGDLM we track Standard Bank and the compare results with those from the DLM of Standard Bank. Using the out-of-sample forecasts, we computed two measures of forecast accuracy, root mean square error (RMSE) and mean absolute deviation (MAD), for each of the stocks, in the SGDLM case and the DLM case, and made comparisons. Table 4 summaries the results. For each stock, we bold the smaller value of the error to indicate the better model.

Table 4: Comparison of measures of forecast accuracy (RMSE and MAD) between the SGDLM and the DLM.

| | **Standard Bank Group** | | | **FirstRand Limited** | |
|---|---|---|---|---|---|
| | SGDLM | DLM | | SGDLM | DLM |
| RMSE | **0.023407** | 0.023458 | RMSE | **0.023118** | 0.023196 |
| MAD | **0.016465** | 0.016469 | MAD | 0.016520 | **0.016498** |
| | **Glencore plc** | | | **Anglo American plc** | |
| | SGDLM | DLM | | SGDLM | DLM |
| RMSE | 0.024470 | **0.024327** | RMSE | 0.025138 | **0.024914** |
| MAD | 0.018218 | **0.018070** | MAD | 0.017994 | **0.017843** |
| | **British American Tobacco** | | | **MTN Group Limited** | |
| | SGDLM | DLM | | SGDLM | DLM |
| RMSE | **0.017620** | 0.017631 | RMSE | 0.031298 | **0.031227** |
| MAD | **0.012833** | 0.012857 | MAD | 0.020034 | **0.019918** |
| | **Naspers** | | | **Shoprite Holdings Limited** | |
| | SGDLM | DLM | | SGDLM | DLM |
| RMSE | 0.026554 | **0.026510** | RMSE | **0.021877** | 0.021982 |
| MAD | **0.018132** | 0.018137 | MAD | **0.015506** | 0.015559 |

According to Table 4, none of the two models outperforms the other in all cases. The SGDLM performs better than the DLM in the case of Standard Bank, British American Tobacco and Shoprite Holdings Limited given that it gives smaller values of both the RMSE and the MAD, but the exact opposite occurs for Glencore plc, Anglo American plc and MTN Group Limited. For FirstRand, the SGDLM produces a smaller value of RMSE than that produced by the DLM, but the exact opposite occurs with the value of MAD; a similar thing occurs with Naspers. For these selected stocks, we observe a tie between the two models. It can be seen that the differences between the errors of the SGDLM and the DLM are very small. The values in Table 4 are run-dependent; they keep changing slightly each time you run the analysis, but the comparison between the two models generally remains the same.

In principle, we expect the SGDLM to outperform the DLM because, unlike the DLM, the SGDLM is a model framework that captures dependencies among stocks. So, for all the stocks, we expected the SGDLM to give more accurate forecasts. This is not the case in the current example. We propose that, to make the SGDLM perform better universally than the DLM, there is a need to improve the formulation of the SGDLM. One aspect here is the selection of simultaneous parents. In the current study, we picked the simultaneous parent of each of the stocks at $t = 782$ and maintained it to the end of the analysis. This is unrealistic because the market is dynamic; a good simultaneous parent to Standard Bank today may not remain good to Standard Bank after, say, one year. So, there is a need to use a method of selecting simultaneous parents that involves refreshing the parents as the analysis proceeds, e.g., Gruber and West (2017).

### 6.3  Comparison of the empirical returns trend with the SGDLM trend

We calculated the empirical/observed 100-day simple moving averages for the returns and compared the resultant trend with that of the SGDLM, for each of the eight stocks. Figure 1 summarises the outcomes.

For any model that fits the data well, the observed trend of the data and the trend of the model forecasts should follow each other closely, if there is no stock market stress. Up to around March 2020 the observed trend and the SGDLM trend follow each other closely for almost all the eight stocks; only British American Tobacco shows a clear discrepancy between the two trends during this period. This discrepancy reflects the up and down movements of the price of British American Tobacco in 2019 (see Figure 2e). The SGDLM overestimates the returns during the market crash that started in March 2020 for all the stocks that were hit hard by the crash. This overestimation is literally visible in the case of Standard Bank, FirstRand, Glencore plc, Anglo American plc and MTN Group Limited, and reflects the profuse drop in the prices in March 2020 (see Figure 2). The observed trend generally trails below the SGDLM trend just after the period of the intense market stress – this is expected because the formula for calculating the 100-moving averages carries along the radically below zero values of the returns for a couple of months after the market crash. Generally, the two trends track each other closely after the impact of the intense market crash.

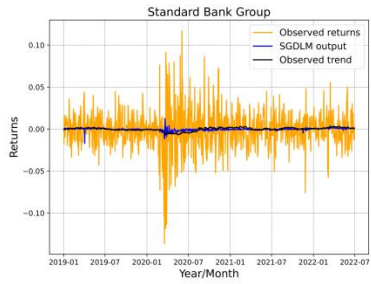

(a)

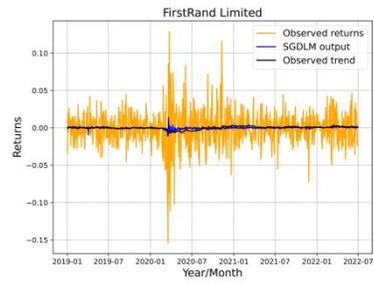

(b)

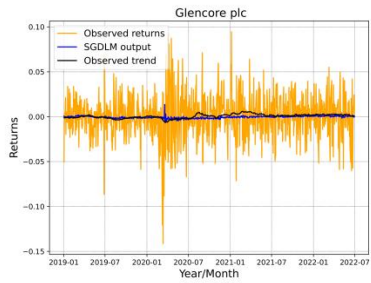

(c)

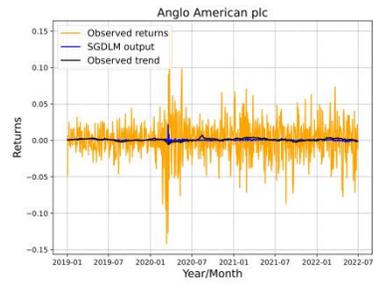

(d)

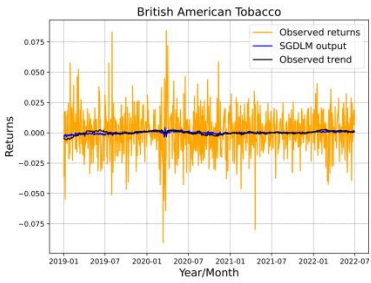

(e)

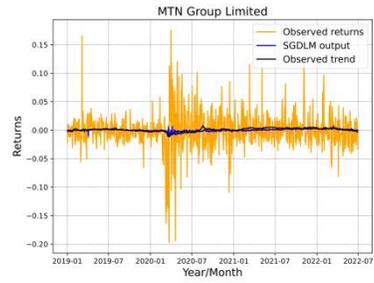

(f)

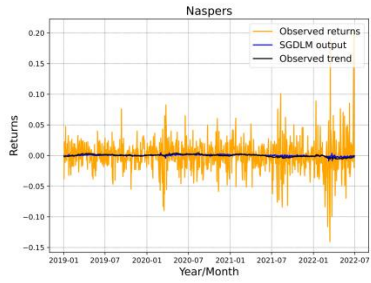

(g)

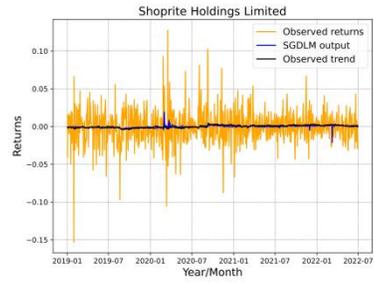

(h)

Figure 1: Comparison of the observed trend of the returns with the SGDLM trend.

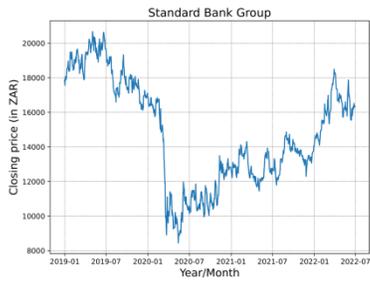

(a)

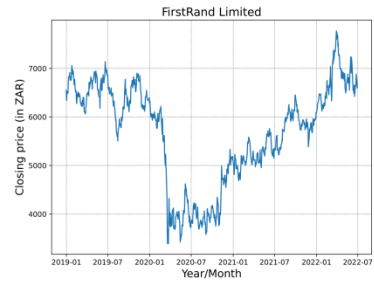

(b)

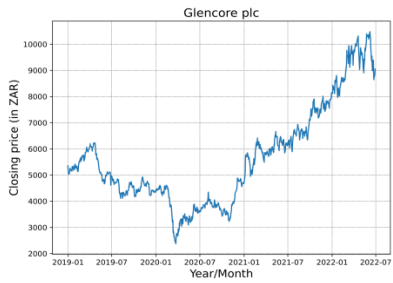

(c)

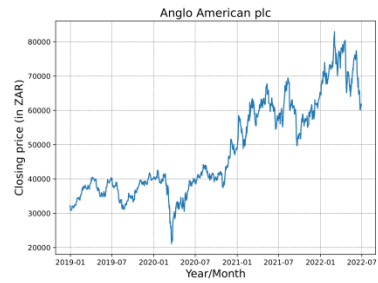

(d)

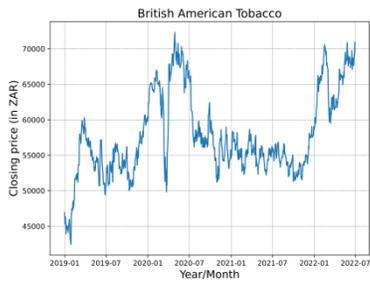

(e)

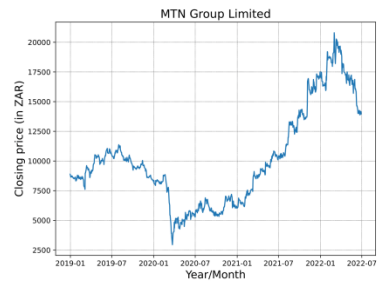

(f)

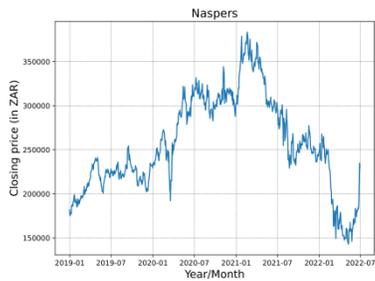

(g)

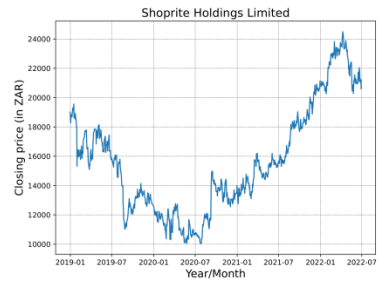

(h)

Figure 2: Closing prices of some stocks over the test data period.

## 6.4 Efficiency of importance sampling and MFVB

In Figure 3, we illustrate the evaluation of the efficiency of the importance sample-based approximation of the exact posterior. The efficiency of the MFVB approach to obtaining the decoupled conjugate forms from the approximated posterior is also evaluated. It can be seen in Figure 3a that for the bigger part of the test period, the effective sample size is above 1,900, which means that the importance sample-based approximation of the posterior is more than 95% effective. The most worrying period starts towards the end of February 2020 up to around mid-April 2020, during which the effective sample size nosedives to 1325 or so (about 66% effective). It should be noted that the first case of COVID-19 was announced in South Africa in early March 2020, and as Figure 2 shows, the result of this announcement was a plunge in the prices of most of the stocks, which subsequently caused a temporary breakdown of the SGDLM and hence the drastic fall in the effective sample size. The other quite radical unexpected fall of the effective sample size is seen in late November 2021 due to the outbreak of the Omicron variant. The SGDLM however recovers from both short-term breakdowns and the importance sample-based posterior approximation is generally good throughout the test period. Correspondingly, Figure 3b shows the KL divergence as a measure of the effectiveness of MFVB. In the SGDLM framework, KL divergence is approximated by the entropy of the importance sample, and because of this, it is expected that whenever ESS is high, KL divergence is low and vice versa. So, the periods when the ESS drops are the very periods when KL divergence goes up. Generally, KL divergence remains small throughout the test period. It is interesting to see that the realised KL divergence does not exceed its theoretical upper bound at any point of the test period.

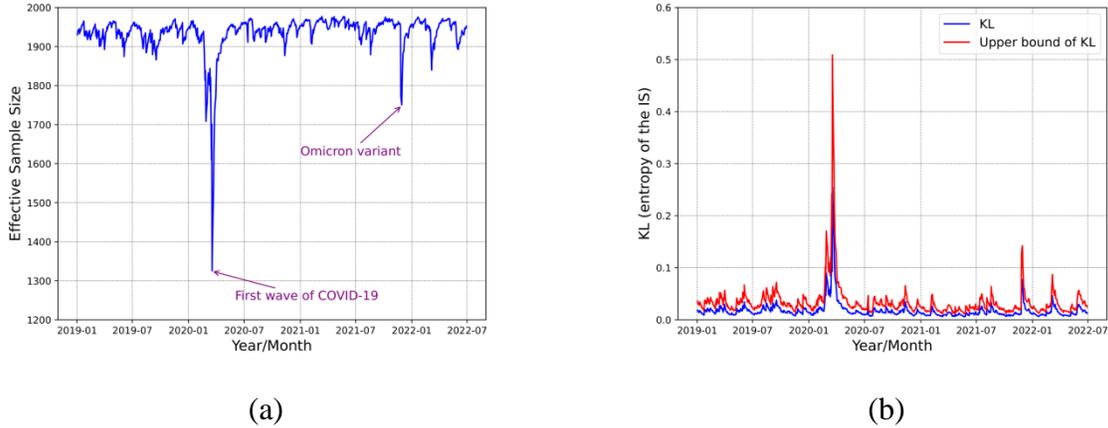

(a)                                     (b)

Figure 3: (a) Measurement of the efficacy of the importance sample (IS) and (b) measurement of the efficacy of the MFVB approximation.

## 6.5 Effect of the number of simultaneous parents on forecast accuracy

We re-ran the full SGDLM analysis with a bigger number of simultaneous parents. We ran the analysis with two and five simultaneous parents. With two simultaneous parents, we found out that the optimal values of discount factors are $\beta = 0.919$, $\delta_\phi = 0.990$, and $\delta_\gamma = 0.970$. With five simultaneous parents, we obtained $\beta = 0.909$, $\delta_\phi = 0.983$, and $\delta_\gamma = 0.984$. In both situations, $K = N = 2{,}000$. In Table 5, we present the aggregate coverage of prediction intervals for the different numbers of simultaneous parents. In Table 6, we compare the RMSE and MAD values for three stocks, for the different parental sizes.

Table 5: Aggregate interval coverage for different parental sizes.

| **Prediction interval (%)** | 99 | 95 | 90 | 80 | 50 | 20 | 10 |
|---|---|---|---|---|---|---|---|
| | **One simultaneous parent** | | | | | | |
| Coverage (%) | 98.4 | 95.7 | 92.6 | 86.0 | 60.4 | 26.5 | 14.4 |
| | **Two simultaneous parents** | | | | | | |
| Coverage (%) | 98.5 | 96.0 | 93.0 | 86.5 | 61.3 | 27.5 | 15.1 |
| | **Five simultaneous parents** | | | | | | |
| Coverage (%) | 98.6 | 96.2 | 93.3 | 87.0 | 62.0 | 28.1 | 15.7 |

Table 6: Comparison of RMSE and MAD across different parental sizes for the SGDLM.

| Standard Bank Group | | | |
|---|---|---|---|
|  | 1 SP | 2 SP | 5 SP |
| RMSE | 0.0234 | 0.0236 | 0.0308 |
| MAD | 0.0165 | 0.0166 | 0.0179 |
| British American Tobacco | | | |
|  | 1 SP | 2 SP | 5 SP |
| RMSE | 0.0176 | 0.0192 | 0.0220 |
| MAD | 0.0128 | 0.0132 | 0.0139 |
| MTN Group Limited | | | |
|  | 1 SP | 2 SP | 5 SP |
| RMSE | 0.0313 | 0.0316 | 0.0334 |
| MAD | 0.0200 | 0.0202 | 0.0209 |

In Table 5, we observe that the SGDLM of one simultaneous parent produces the most concise prediction intervals, followed by the one with two, and the one of five comes last. The percentages in the table are the averages across all stocks; the percentages for individual stocks across the different parental sizes are expected to be similar to those in the aggregate. In Table 6, the SGDLM of one simultaneous parent gives the smallest errors. The errors produced by the SGDLM of two simultaneous parents are bigger than those of the SGDLM of one simultaneous parent but smaller than those of the SGDLM of five simultaneous parents. Results from both tables suggest that the SGDLM with one simultaneous parent is the most accurate, followed by the one with two, and the one of five comes last. This supports our use of one simultaneous parent in the analyses of the preceding sections. Thus, with the current dimension of 40 stocks, using one simultaneous parent produces the most accurate results. The results of the table also suggest that accuracy reduces as the number of simultaneous parents increases. However, it should be noted that, in higher dimensions, the most accurate results may be obtained when using more than one simultaneous parent, e.g., Gruber and West (2016, 2017).

# 7 Conclusion

Our study has shown that the SGDLM forecasts the returns of the stock data accurately. With a dimension of 40 stocks or less, our results suggest that the most accurate forecasts are obtained with one simultaneous parent. Our insights into the efficiency of the recoupling/decoupling techniques indicate that the techniques perform well generally and that the SGDLM responds well to changes in the market. As a possible direction for future research, by adopting the approach of refreshing simultaneous parents depending on the prevailing market conditions, e.g., Gruber and West (2017), we recommend re-doing the comparison of RMSE and MAD between the DLM and the SGDLM. This will potentially make the SGDLM to outperform the DLM universally.

**Funding**

The first author received funding from Stellenbosch University. Opinions expressed and conclusions arrived at, are those of the authors and are not necessarily to be attributed to Stellenbosch University.

**Appendix A: The JSE stocks used**

Table 7: The selected 40 JSE companies and their sector categorisations.

| Financials | Basic materials |
|---|---|
| FirstRand Limited | Glencore plc |
| Standard Bank Group | Anglo American plc |
| Capitec Bank Holdings | Anglo American Platinum Limited |
| ABSA Group Limited | Sasol Limited |
| Nedbank Group Limited | Kumba Iron Ore |
| Discovery Limited | Impala Platinum Holdings Limited |
| Remgro Limited | AngloGold Ashanti |
| PSG Group Limited | Exxaro Resources Limited |
| Nepi Rockcastle plc | African Rainbow Minerals |
| Santam Limited | Sappi Limited |
| Transaction Capital Limited | |
| Investec Limited | |

| Consumer services | Consumer goods |
|---|---|
| Shoprite Holdings Limited | British American Tobacco |
| Clicks Group Limited | Compagnie Fin Richemont |
| Woolworths Holdings Limited | Tiger Brands Limited |
| Mr Price Group | AVI Limited |
| Pick n Pay Stores Limited | |
| Spar Group Limited | |
| Truworths International Limited | |

| Telecommunications | Industrials |
|---|---|
| MTN Group Limited | Bidvest Group |
| Vodacom Group Limited | Barloworld Limited |
| Telkom SA Limited | |

| Technology | Health care |
|---|---|
| Naspers | Aspen Pharmacare Holdings Limited |

**Appendix B: Computation time**

The runtimes for the SGDLM analysis, for the different parental sizes, are presented here. We did all analyses using a 2017 desktop computer with a CPU of 3.20 GHz, four cores, and 8 GB RAM. In all analyses, $K = N = 2{,}000$. Phase 1 of the SGDLM implementation took about 9 seconds. However, phases 2 and 3 had much longer runtimes. Table 8 shows the approximate number of hours taken for the analysis to execute. It should be noted that we took less time than what is shown in the table because we could run three Jupyter Notebooks at once to select the discount factors; for example, for the analysis that involves using one simultaneous parent, the total runtime for phases 2 and 3 was $19 + 4 + 14 = 37$ hours. This computation time is much higher than that realised when using GPU-accelerated computing, e.g., Gruber and West (2016).

Table 8: Runtime (in hours) of the SGDLM implementation for the different parental sizes.

|  |  | 1 SP | 2 SP | 5 SP |
|---|---|---|---|---|
| Phase 2 | Selection of discount factors | $19 \times 3$ | $23 \times 3$ | $29 \times 3$ |
|  | Obtaining initial priors | 4 | 5 | 6 |
| Phase 3 |  | 14 | 15 | 19 |
| **Total** |  | **75** | **89** | **112** |

**Appendix C: Formulae for mean-field posterior**

By denoting $E[\cdot]$ as the expectation with respect to the importance sample probability measure $p_{\text{MC}}$, the formulae (whose proofs can be found in Kyakutwika (2022), Xie (2021), and Gruber (2015)) for obtaining the parameters of the mean-field variational Bayes posteriors are as follows. Firstly, compute:

$$\begin{aligned} \mathbf{m}_{it} &= \frac{E[\lambda_{it}\boldsymbol{\theta}_{it}]}{E[\lambda_{it}]} \\ \mathbf{V}_{it} &= E[\lambda_{it}(\boldsymbol{\theta}_{it} - \mathbf{m}_{it})(\boldsymbol{\theta}_{it} - \mathbf{m}_{it})^T] \\ d_{it} &= E[\lambda_{it}(\boldsymbol{\theta}_{it} - \mathbf{m}_{it})^T \mathbf{V}_{it}^{-1}(\boldsymbol{\theta}_{it} - \mathbf{m}_{it})] \end{aligned}$$

Then, $n_{it}$ is the unique solution to

$$\ln(n_{it} + p_i - d_{it}) - \psi\left(\frac{n_{it}}{2}\right) - \frac{(p_i - d_{it})}{n_{it}} - \ln(2E[\lambda_{it}]) + E[\lambda_{it}] = 0$$

where $\psi$ is the digamma function. Finally, we set:

$$\begin{aligned} s_{it} &= \frac{n_{it} + p_i - d_{it})}{n_{it} E[\lambda_{it}]} \\ \mathbf{C}_{it} &= s_{it} \mathbf{V}_{it} \end{aligned}$$

**Appendix D: Our Python code for the SGDLM implementation**

All Python codes for the analyses in this study are available on the GitHub page:
github.com/nelsonkyakutwika/SGDLM